\documentclass{JHEP3} 

\usepackage{epsfig,multicol}
\usepackage{amsmath, amssymb}
\usepackage{epic}
\usepackage{concmath, palatino}

\newcommand\fverb{\setbox\pippobox=\hbox\bgroup\verb}
\newcommand\fverbdo{\egroup\medskip\noindent%
			\fbox{\unhbox\pippobox}\ }
\newcommand\fverbit{\egroup\item[\fbox{\unhbox\pippobox}]}
\newbox\pippobox

\newcommand{\be}{\begin{equation}} 
\newcommand{\ee}{\end{equation}}

\newcommand{\ba}{\begin{eqnarray}}
\newcommand{\ea}{\end{eqnarray}}

\newcommand{\ads}{AdS_5\times S^5}

\title{Anomalous dimensions of finite size field strength operators in ${\cal N}=4$ SYM}

\author{Matteo Beccaria\\
  Dipartimento di Fisica, Universita' del Salento,
  Via Arnesano, 73100 Lecce\\
  INFN, Sezione di Lecce\\
  E-mail: \email{matteo.beccaria@le.infn.it}}

\author{Valentina Forini\\
  Humboldt-Universit\"{a}t zu Berlin, Institut f\"{u}r Physik, Newtonstra{\ss}e
  15, D-12489 Berlin\\and\\
  Dipartimento di Fisica, Universita' di Perugia,
  Via A. Pascoli, I-06123 Perugia\\
  INFN, Sezione di Perugia\\ 
  E-mail: \email{forini@physik.hu-berlin.de,forini@pg.infn.it}}

\preprint{HU-EP-07/29}

\abstract{
In the ${\cal N}=4$ super Yang-Mills theory, we consider the higher order anomalous dimensions $\gamma_L(g)$ of purely gluonic operators
$\mbox{Tr}\,{\cal F}^L$ where $\cal F$ is a component of the self-dual field strength. We propose compact closed expressions depending parametrically 
on $L$ that reproduce the prediction of Bethe Ansatz equations up to five loop order, including transcendental dressing corrections. The size dependence follows a simple
pattern as the perturbative order is increased and suggests hidden relations for these special operators. 
}

\begin{document} 

\section{Introduction}
\label{sec:Intro}

Integrable structures emerge as a deep property of four dimensional 
Yang-Mills theories in the 't Hooft planar limit. In the simplest context, integrability 
underlies and governs the scale evolution of renormalized 
composite operators belonging to specific subsectors of the theory~\cite{Belitsky:2004cz}.

Historically, this intriguing phenomenon was discovered in the study of planar QCD, definitely a non-trivial quantum 
theory~\cite{QCD-int}. At one-loop, suitable maximal helicity Wilson operators admit a peculiar  
renormalization mixing matrix, the dilatation operator. It can be identified with the Hamiltonian of
integrable $XXX$ spin chains with $\mathfrak{sl}(2, \mathbb{R})$ symmetry. 
This is a light-cone subalgebra of the full four dimensional 
conformal algebra $\mathfrak{so}(4,2)$.

>From a modern perspective, conformal symmetry, unbroken in QCD at one-loop, does not appear to be a necessary condition for 
integrability, as discussed in~\cite{Belitsky:2004sc,Belitsky:2004sf,Belitsky:2005bu,DiVecchia:2004jw}.
Nevertheless, it plays an important role by imposing selection rules and multiplet structures 
and is helpful to clarify the origin and details of integrability.
The same reasoning applies to supersymmetric extensions of QCD with ${\cal N}=1,2,4$ supercharges. In particular,  
multiplets of composite operators are greatly simplified in the maximal superconformal ${\cal N}=4$ theory~\cite{Belitsky:2005gr}.
Also, intermediate level integrability is achieved in various orbifold reductions of ${\cal N}=4$ SYM reducing the number 
of supercharges~\cite{Wang:2003cu}.

As is well known, in the maximal ${\cal N}=4$ case another conceptual tool is available to 
deepen the investigation, namely Maldacena AdS/CFT duality~\cite{Maldacena:1997re}~\footnote{For those aspects of the duality that most concern our analysis, we 
refer the reader to~\cite{Beisert:2004ry,Plefka:2005bk}.}. It relates ${\cal N}=4$ SYM and $\ads$ superstring which is classically
integrable~\cite{Bena:2003wd}.
Currently, a lot is known about the duality between the integrability properties of the two 
sides of the correspondence with a continuous very stimulating back and forth feeding.
In particular, AdS/CFT duality has been a crucial ingredient to arrive at the  
higher loop proposal for the $S$-matrix of ${\cal N}=4$ 
SYM~\cite{Arutyunov:2004vx,Staudacher:2004tk,Park:2005ji,Beisert:2005fw,Beisert:2005tm,Janik:2006dc,Eden:2006rx,Hernandez:2006tk,Freyhult:2006vr,Plefka:2006ze,Beisert:2006ib,Beisert:2006ez}.

Forgetting for a while the string side, we can ask what kind of understanding can be gained from 
integrability in the gauge theory. This is a natural  issue if we are ultimately interested in low-energy physical 
applications to hadronic phenomenology. As a first step, we can honestly postpone the important problem of 
taming conformal and supersymmetry breaking and the precise link with QCD~\cite{Dokshitzer:2006nm}. Working within the ${\cal N}=4$ SYM theory, we can 
examine the outcomes and limitations  of integrability as far as it is currently understood.

>From this point of view, integrability can be regarded as a tool for multi-loop computations,
although this attitude could be admittedly narrow-minded. 
The typical object that is computed are higher order corrections to the anomalous dimension
of specific composite operators. In all cases, these operators are single traces of the general form 
\be
{\cal O} = \mbox{Tr}\,\left(\,\prod_{i=1}^L D^{n_i}\, X_i\right) + \mbox{permutations},
\ee
where $X_i$ are elementary fields in certain subsectors of the full ${\cal N}=4$ SYM and covariant derivatives
generically appear to close the renormalization mixing.

In the most favorable cases, we are able to write down Bethe Ansatz equations providing the anomalous dimension of ${\cal O}$
as a perturbative series in the  't Hooft coupling~$g$
\be
\label{eq:gamma}
\gamma_{\cal O}(g) = \sum_{n\ge 0} c_n({\cal O})\, g^{2\,n}. 
\ee
Such formidable results face a first and major limitation, namely 
the well-known {\em wrapping problem} (see~\cite{Kotikov:2007cy,Janik:2007wt} for recent developments). The coefficients $c_n$ are reliable up to a maximum order 
$n_{\rm max}$
that typically depends linearly on $L$. This means that $\gamma_{\cal O}(g)$ is actually calculable up to, say, ${\cal O}(g^{2L})$
terms - a stumbling wall to any extrapolation to the genuine strong coupling regime.
A notable exception occurs in the $L\to\infty$ thermodynamical limit. Then, wrapping is absent and resummations of Eq.~(\ref{eq:gamma}) can be attempted 
to match string duality 
predictions~\cite{Beisert:2006ez,Rej:2007vm}.

A more subtle limitation appears when we try to investigate the dependence on $L$ and $\{n_i\}$ at fixed perturbative order. Apart from very special 
cases, the Bethe equations do not provide the expansion coefficients as functions of 
$L$ and $\{n_i\}$, but just provide sequences of numerical (sometimes rational) values for each given operator.
This is an unwanted situation as can be appreciated in the sector of the so-called twist operators~\cite{Lipatov:1997vu}. These are operators with a certain 
phenomenological origin in the QCD case. The length $L$ is fixed and one would like to know the analytic dependence of $\gamma(g, S)$
on the spin quantum number $S = \sum_i n_i$. For instance, this is a standard procedure to analyze BFKL physics of pomeron exchange~\cite{Lipatov:1976zz}.

Very recently, intense work on twist-2 and 3 operators has led to higher order conjectures
for the functions $c_n(S)$ appearing in the expansion $\gamma(g, S) = \sum_{n} c_n(S)\,g^{2\,n}$~\cite{Kotikov:2007cy,twist}. Proofs are however missing, at least beyond one-loop. It seems that new tools
are needed  to derive them rigorously from the Bethe Ansatz equations.

If we give up exact results and turn to approximation, systematic methods can be applied to extract 
the large $L$, $n_i$ corrections. Indeed, this is a thermodynamical limit of the underlying spin chain
where both the length and the number of magnons grows to infinity. Various techniques are available and 
have been successfully applied to rank-1 $\mathfrak{su}(2)$ and $\mathfrak{sl}(2)$ 
subsectors~\cite{Lubcke:2004dg,Beisert:2005bv,Gromov:2005gp,Feverati:2006hh,Casteill:2007ct}.
For higher rank sectors (see also~\cite{Freyhult:2004iq,Freyhult:2005fn,Freyhult:2006vr} for an analysis in the rank-2  $\mathfrak{su}(3)$ sector) the techniques developed in~\cite{deVega:1987vh} could be useful. 
Indeed, in the recent~\cite{Gromov:2007ky} an integral equation describing
finite size corrections to the full nested Bethe ansatz was derived.

On the string side of the AdS/CFT correspondence, there is an analogously intense ongoing discussion on how finite size effects of the 
string world-sheet could modify the solvability of the string sigma-model in AdS$_5\times S^5$ by means of a 
Bethe ansatz~\cite{Kazakov:2004qf,Arutyunov:2004vx,Staudacher:2004tk,Beisert:2005bm,Beisert:2005fw,Beisert:2005cw}. The currently known Bethe equations for quantum strings in AdS$_5\times$S$_5$ are 
asymptotic and describe the string spectrum with an exponential accuracy as long as the string length is sufficiently 
large~\cite{SchaferNameki:2006ey}.
The breakdown of the asymptotic approximation via exponential terms, firstly described from a field theory point of view in~\cite{Ambjorn:2005wa}, has been determined for 
the spectrum of spinning strings in the $\mathfrak{su}(2)$ and $\mathfrak{sl}(2)$ sector~\cite{SchaferNameki:2006gk,SchaferNameki:2006ey} and for the 
giant magnon~\cite{Hofman:2006xt} dispersion relation in~\cite{Arutyunov:2006gs,Astolfi:2007uz}. 
In particular, the exponential term in the finite size correction to the giant magnon dispersion relation
has been recently and nicely rederived in~\cite{Janik:2007wt} via a generalization of known results in relativistic quantum field theory, and there is a 
general and deep interest in obtaining exact results which should be valid for any value of the string lenght, which is in turn proportional to the lenght ($R$-charge)
of the corresponding gauge operator.

In this paper, we contribute to the above general discussion and consider, in the gauge field theory context, a special class of operators
where finite size corrections can be given in closed form. In other words, 
we provide the coefficients $c_n$ in Eq.~(\ref{eq:gamma}) as exact functions of the operator length $L$.
This is a seemingly unique result which, although peculiar, is very interesting and puzzling and certainly deserves some attention.

The considered operators have a complicated mixing pattern and reduce at one loop to the  purely gluonic higher dimensional 
condensates of the form 
\be
{\cal O}_L = \mbox{Tr}\,{\cal F}^L,
\ee
where ${\cal F}$ is one component of the self-dual Yang-Mills field strength. The operators ${\cal O}_L$ are exact eigenstates 
of the one-loop dilatation operator and can be mapped to the ferromagnetic states of an integrable
spin $S=1$ chain~\cite{Ferretti:2004ba,Beisert:2004fv}. 
As such, their one-loop dependence on the length $L$ is trivial and (including the classical dimension), it is known 
that~\cite{Ferretti:2004ba,Beisert:2004fv,SchaferNameki:2004ik,Rej:2007vm}
\be
\gamma_{{\cal O}_L} = L\,(2 + 3\,g^2 + \cdots),
\ee
Beyond one-loop, the analysis of~\cite{Rej:2007vm} provide efficient computational tools to derive the sequence $\{c_n(L)\}$ 
for any given $L$, although not parametrically. It turns out immediately that $c_n(L)$ is not linear in $L$
as far as $n\ge 2$. So, starting at two-loops, non-trivial finite size corrections appear.

In this paper, we  analyze the sequences $\{c_n(L)\}$ at fixed $n$ as $L$ is varied up to large values. By a careful investigation 
of the (infinite precision) numerics, we conjecture and provide closed expressions for $c_n(L)$ valid up to 5 loops, including the 
transcendental terms coming from the $S$-matrix dressing phase~\cite{Beisert:2005tm}. To give an example, the two-loop
anomalous dimension takes the remarkably simple form 
\be
\gamma_L(g) = 2\,L+3\,L\,g^2 + \left(-\frac{51}{8}+\frac{9}{8}\,\frac{1}{(-1)^L\,2^{L-1}+1}\right)\,L\,g^4 + \cdots\, ,
\ee
with exponentially suppressed corrections to the trivial linear scaling with $L$. We have been able to extend the above equation up to 
five loops. The detailed results will be illustrated in the main text. Here, we 
just anticipate the large $L$ limit which reads
\be
\frac{\gamma_L(g)}{L} = f_0(g) + g^4\, h(g\,L)\, e^{-L\,\log\,2} + {\cal O}(e^{-2\,L\,\log\,2}), 
\ee
where $f_0(g)$ has been computed in~\cite{Rej:2007vm}. The function $h(z)$ is regular around $z=0$ and does not receive contributions from the dressing phase, 
at least up to five loops. The size corrections to the thermodynamical limit are thus characterized by a finite specific
correlation length $\xi = 1/\log\,2$. In the final Section of the paper, we shall try to argue why this correlation length abruptly appears at two-loops breaking the 
trivial linear dependence on $L$.

\section{One-loop ferromagnetic multi-gluon operators in the chiral sector}
\label{sec:oneloop}

In this Section, we introduce the special class of ${\cal N} = 4$ multi-gluon operators that we are going to analyze.
For completeness, we also review their one-loop integrability properties and, in particular, the reduction of the mixing matrix to the 
Hamiltonian of an integrable $XXX_1$ chain.

In the planar limit, the most general purely gluonic local gauge invariant operators are easily identified.
They are single trace operators built with covariant derivatives of the field strength
\be
\label{eq:gluonic}
\mbox{Tr}\left(D^{n_1} F_{\mu_1\nu_1}\cdots D^{n_L} F_{\mu_L\nu_L}\right).
\ee
The anomalous dimension matrix $\Gamma$ and the would-be spin chain Hamiltonian $H$ are related by 
\be
\Gamma = \mu\frac{\partial}{\partial\mu}\,\log\,Z \equiv g^2\,H,\qquad g^2 = \frac{g_{\rm YM}^2\,N_c}{8\pi^2},
\ee
where $Z$ is the renormalization matrix and $g$ is the scaled 't Hooft coupling kept fixed in the planar limit $N_c\to\infty$.

The one-loop operator $H$ takes the form of a nearest neighbor Hamiltonian conserving the length $L$ in Eq.~(\ref{eq:gluonic}). It can be written
\be
H = \sum_{n=1}^L H_{n,n+1},
\ee
where the link Hamiltonian $H_{n,n+1}$ acts on the fields at positions $n$ and $n+1$ and is independent on $n$.
For the complete ${\cal N} = 4$ SYM theory, the elementary fields are included in the singleton multiplet $V$~\cite{Beisert:2004ry} and 
the link Hamiltonian reads~\cite{Beisert:2003jj}
\be
H^{{\cal N}=4} = 2\,\sum_{j=0}^\infty h(j)\,P_j^{{\cal N}=4},\qquad h(j) = \sum_{n=1}^j\frac{1}{n},
\ee
where  $P_j^{{\cal N}=4}$ is a projector onto the irreducible superconformal multiplets appearing in the decomposition of the 
two-site states $V\otimes V$. 

To restrict the analysis to purely gluonic operators it is convenient to adopt the conformal analysis
exploited in the QCD reduction described in~\cite{Ferretti:2004ba,Beisert:2004fv} (see 
also~\cite{Gorsky:2004kb}). We first split into irreducible components 
the field strength $F_{\mu\nu}$ transforming as $(1,0)\oplus(0,1)$ under the 
$\mathfrak{so}(3,1)=\mathfrak{su}(2)\oplus\mathfrak{su}(2)$ Lorentz algebra. This is achieved by means of the 't Hooft symbols~\cite{'tHooft:1976fv}
projecting $F_{\mu\nu}$ onto the self-dual $(1,0)$ and anti-self-dual $(0,1)$ components
\be
F_{\mu\nu} = \eta^A_{\mu\nu}\,f^A + \overline{\eta}^A_{\mu\nu}\,\overline{f}^A,\qquad A = 1, 2, 3.
\ee
The purely chiral gluon operators are the subset of Eq.~(\ref{eq:gluonic}) built using only 
the self-dual part of $F_{\mu\nu}$ 
\be
\mbox{Tr}\left\{D^{n_1} f^{A_1}\cdots D^{n_L} f^{A_L}\right\}.
\ee
At one loop, they close under renormalization mixing. The relevant link Hamiltonian can be obtained by
restriction of $H^{{\cal N}=4}$. To this aim, it is convenient to organize the various covariant derivatives of $f^A$
in a ${\cal N}=0$ conformal infinite dimensional multiplet
\be
V^f = \{D^n\,f\}_{n\ge 0}.
\ee
Two-site states decompose in irreducible multiplets labeled by the conformal spin $j$ according to 
\be
\label{eq:tensor}
V^f\otimes V^f = \bigoplus_{j=-2}^\infty V_j^{ff}.
\ee
Also, the conformal splitting of the full ${\cal N}=4$ projector $P_j^{{\cal N}=4}$ turns out to involve 
the conformal projector $P_{j-2}^{ff}$ only. This leads to the following purely gluonic link Hamiltonian in the chiral sector
\be
\label{eq:tmp1}
H = 2\,\sum_{j=-2}^\infty h(j+2)\,P_j^{ff}.
\ee
Finally, if we further restrict to operators without derivatives, one can prove that the only modules appearing 
in the r.h.s. of Eq.~(\ref{eq:tensor}) are those with $j=-2, -1, 0$~\cite{Beisert:2004fv}.  To make contact with the spin chain interpretation, we
introduce spin $S=1$ $\mathfrak{su}(2)$ operators $\{S^i\}$ acting on the three components $f^A$ as
\be
(S^i\,f)^A = i\,\varepsilon^{iAB}\,f^B.
\ee
Then, the modules $P^{ff}_j$ with $j=-2,-1,0$ can be shown to be associated with the $\mathfrak{su}(2)$ representations 
with $S=0, 1, 2$ respectively, appearing in the decomposition $1\otimes 1 = 2\oplus 1\oplus 0$. The link Hamiltonian $H_{n,n+1}$ 
in Eq.~(\ref{eq:tmp1}) can be written as a polynomial in $\mathbf{S}_n\cdot \mathbf{S}_{n+1}$ with the result
\be
H = 3L + \frac{1}{2}\sum_n \left[\mathbf{S}_n\cdot \mathbf{S}_{n+1}-(\mathbf{S}_n\cdot \mathbf{S}_{n+1})^2\right].
\ee
This is an anti-ferromagnetic integrable spin-chain that can be diagonalized by Bethe Ansatz~\cite{spin1}.
The ground state is highly non-trivial, but the maximally excited states are a trivial ferromagnetic
multiplet. A convenient representative is the operator
\be
{\cal O}_L = \mbox{Tr}({\cal F}^L),
\ee 
where ${\cal F} = f^+$ is the maximal eigenstate of $S^z$. The anomalous dimension of this state (including the classical dimension) is simply
\be
\label{eq:oneloopgamma}
\gamma_L(g) = 2\,L + 3\, L\, g^2 + {\cal O}(g^2).
\ee
The linear dependence on $L$ follows from the uniform structure of the ferromagnetic state.

In this paper, we shall be working on the conformal/field-theory side of the AdS/CFT correspondence. However, it must be mentioned that the natural candidate for a semiclassical string state dual to ${\cal O}_L$ has been proposed in~\cite{Frolov:2003qc,Park:2005kt}. It describes a rigid circular string rotating simultaneously in two orthogonal spatial planes of $AdS_5$ with equal spins $S \equiv L$. At large $S$, the weak-coupling extrapolation for the energy is given by
\ba\nonumber
E &=& p(\lambda)\,S+q(\lambda)+\dots,\\
p_{\lambda\gg1}&=&p_0+\frac{p_1}{\sqrt{\lambda}}+\dots\, ,\quad\quad\quad q_{\lambda\gg1}=\sqrt{\lambda}q_0+q_1+\dots\, .
\label{endual}
\ea
The results for the three-level and 1-loop coefficients of the solution have been calculated in~\cite{Park:2005kt}  within a stability region $0.4\lesssim\frac{S}{\sqrt{\lambda}}\gtrsim 1.17$, corresponding to a fixed value ($m=1$) of the winding number. It should be noticed that, since in the semiclassical approximation $\lambda$ is large on the string side, the interval of stability for the solution does include large values of $S$,  allowing the comparison to large $S$, large $\lambda$ asymptotics of the exact anomalous dimension. The linear dependence on $S$ exhibited by the solution (\ref{endual}) supports  the identification of the gauge theory operator  ${\cal O}_L$ with this particular rigid spinning string solution.

\section{Higher loop extension of the scaling field ${\cal O}_L$}

At more than one-loop, the operator ${\cal O}_L$ ceases to be an eigenstate of the dilatation
operator because  the purely gluonic chiral sector does not close under mixing anymore. The higher order 
scaling operator receives corrections and contributions from the other sectors of the full $\mathfrak{psu}(2,2|4)$ theory
and is uniquely defined by the boundary condition of being ${\cal O}_L$ at one-loop.
In the following, we shall not be pedantic about this distinction and keep naming ${\cal O}_L$ the multi-loop
extension of $\mbox{Tr}{\cal F}^L$.

A great deal of information about ${\cal O}_L$ has been obtained in~\cite{Rej:2007vm} in the framework of the long-range Bethe Ansatz equations.
In this Section, we quickly summarize these results with some additional investigation of the finite but large $L$ Bethe roots.
This will fix the setup for the computation of $\gamma_L(g)$.

\subsection{Dynkin diagrams and Bethe roots}

As is well known, several choices are available for the Dynkin diagram of a Lie superalgebra. 
In the case of $\mathfrak{psu}(2,2|4)$, the one loop analysis of 
the operators ${\cal O}_L$ is almost trivial with the Kac distinguished form.
Indeed, ${\cal O}_L$ is the vacuum state and no calculation is needed. 

On the other hand, the all-loop Bethe equations are known for a limited set of (different) choices of the 
Dynkin diagram~\cite{Beisert:2005fw}. In particular, we shall work with the following one
\be
\begin{minipage}{260pt}
\setlength{\unitlength}{1pt}
\small\thicklines
\begin{picture}(260,55)(-10,-30)
%
\put(-32,0){\line(1,0){22}}  
\put(  0,00){\circle{15}}
\put( -5,-5){\line(1, 1){10}}  
\put( -5, 5){\line(1,-1){10}}  
\dottedline{3}(8,0)(32,0)    
\put( 40,00){\circle{15}}     
\dottedline{3}(48,0)(72,0)   
\put( 80,00){\circle{15}}
\put( 75,-5){\line(1, 1){10}}  
\put( 75, 5){\line(1,-1){10}}  
\put( 87,00){\line(1,0){26}} 
\put(120,00){\circle{15}}
\put(120,15){\makebox(0,0)[b]{$+1$}} 
\put(127,00){\line(1,0){26}} 
\put(160,00){\circle{15}}
\put(155,-5){\line(1, 1){10}}  
\put(155, 5){\line(1,-1){10}}  
\dottedline{3}(168,0)(192,0) 
\put(200,00){\circle{15}}
\dottedline{3}(208,0)(232,0) 
\put(240,00){\circle{15}}
\put(235,-5){\line(1, 1){10}} 
\put(235, 5){\line(1,-1){10}} 
\put(250,0){\line(1,0){20}} 
\end{picture}
\end{minipage}
\ee
With respect to this Dynkin diagram, the vacuum is the BPS state $\mbox{Tr}\, \mathcal{Z}^L$ and ${\cal O}_L$
is a highly excited state with many excitations, whose momenta have to be diagonalized by solving the Bethe 
Ansatz equations in order to reproduce the correct energy.
The excitation pattern of Bethe roots for ${\cal O}_L$ is 
\be
\left(K_1,K_2,K_3,K_4,K_5,K_6,K_7\right)=\left(0,0,2L-3,2L-2,L-1,L-2,L-3\right)
\ee
where $K_i$ is the excitation number of the $i$-th node
of the Dynkin diagram
\be
\begin{minipage}{260pt}
\setlength{\unitlength}{1pt}
\small\thicklines
\begin{picture}(260,55)(-10,-30)
%
\put(-32,0){\line(1,0){22}}  
\put(  0,00){\circle{15}}
\put( -5,-5){\line(1, 1){10}}  
\put( -5, 5){\line(1,-1){10}}  
\dottedline{3}(8,0)(32,0)    
\put( 40,00){\circle{15}}     
\dottedline{3}(48,0)(72,0)   
\put( 80,00){\circle{15}}
\put( 75,-5){\line(1, 1){10}}  
\put( 75, 5){\line(1,-1){10}}  
\put( 80,-15){\makebox(0,0)[t]{$2L-3$}}  
\put( 87,00){\line(1,0){26}} 
\put(120,00){\circle{15}}
\put(120,15){\makebox(0,0)[b]{$+1$}} 
\put(120,-15){\makebox(0,0)[t]{$2L-2$}} 
\put(127,00){\line(1,0){26}} 
\put(160,00){\circle{15}}
\put(155,-5){\line(1, 1){10}}  
\put(155, 5){\line(1,-1){10}}  
\put(160,-15){\makebox(0,0)[t]{$L-1$}} 
\dottedline{3}(168,0)(192,0) 
\put(200,00){\circle{15}}
\put(200,-15){\makebox(0,0)[t]{$L-2$}} 
\dottedline{3}(208,0)(232,0) 
\put(240,00){\circle{15}}
\put(235,-5){\line(1, 1){10}} 
\put(235, 5){\line(1,-1){10}} 
\put(240,-15){\makebox(0,0)[t]{$L-3$}} 
\put(250,0){\line(1,0){20}} 
\end{picture}
\end{minipage}
\ee
All but the first two nodes are highly excited.

All the one-loop Bethe equations can be exhibited as roots of explicit polynomial by means of the dualization procedure illustrated in~\cite{Beisert:2005di}
to which we defer the reader for more details.
It is instructive to describe the procedure in graphical terms. Dualizing first at nodes 3 and 7, we obtain 
\be
\begin{minipage}{260pt}
\setlength{\unitlength}{1pt}
\small\thicklines
\begin{picture}(260,55)(-10,-30)
%
\put(-32,0){\line(1,0){22}}  
\put(  0,00){\circle{15}}
\put( -5,-5){\line(1, 1){10}}  
\put( -5, 5){\line(1,-1){10}}  
\dottedline{3}(8,0)(32,0)    
\put( 40,00){\circle{15}}     
\put( 35,-5){\line(1, 1){10}}  
\put( 35, 5){\line(1,-1){10}}  
\put( 47,00){\line(1,0){26}} 
\put( 80,00){\circle{15}}
\put( 75,-5){\line(1, 1){10}}  
\put( 75, 5){\line(1,-1){10}}  
\dottedline{3}(88,0)(112,0)  
\put(120,00){\circle{15}}
\put(115,-5){\line(1, 1){10}}  
\put(115, 5){\line(1,-1){10}}  
\put(120,15){\makebox(0,0)[b]{$+1$}} 
\put(120,-15){\makebox(0,0)[t]{$2L-2$}} 
\put(127,00){\line(1,0){26}} 
\put(160,00){\circle{15}}
\put(155,-5){\line(1, 1){10}}  
\put(155, 5){\line(1,-1){10}}  
\put(160,-15){\makebox(0,0)[t]{$L-1$}} 
\dottedline{3}(168,0)(192,0) 
\put(200,00){\circle{15}}
\put(195,-5){\line(1, 1){10}} 
\put(195, 5){\line(1,-1){10}} 
\put(200,-15){\makebox(0,0)[t]{$L-2$}} 
\put(207,00){\line(1,0){26}} 
\put(240,00){\circle{15}}
\put(235,-5){\line(1, 1){10}} 
\put(235, 5){\line(1,-1){10}} 
\dottedline{3}(250,0)(270,0) 
%
\end{picture}
\end{minipage}
\ee
Dualizing at nodes 4 and 6, we obtain
\be
\begin{minipage}{260pt}
\setlength{\unitlength}{1pt}
\small\thicklines
\begin{picture}(260,55)(-10,-30)
%
\put(-32,0){\line(1,0){22}}  
\put(  0,00){\circle{15}}
\put( -5,-5){\line(1, 1){10}}  
\put( -5, 5){\line(1,-1){10}}  
\dottedline{3}(8,0)(32,0)    
\put( 40,00){\circle{15}}     
\put( 35,-5){\line(1, 1){10}}  
\put( 35, 5){\line(1,-1){10}}  
\put( 47,00){\line(1,0){26}} 
\put( 80,00){\circle{15}}
\put( 87,00){\line(1,0){26}} 
\put(120,00){\circle{15}}
\put(115,-5){\line(1, 1){10}}  
\put(115, 5){\line(1,-1){10}}  
\put(120,15){\makebox(0,0)[b]{$-1$}} 
\dottedline{3}(128,0)(152,0) 
\put(160,00){\circle{15}}
\put(155,-5){\line(1, 1){10}}  
\put(155, 5){\line(1,-1){10}}  
\put(160,15){\makebox(0,0)[b]{$+2$}} 
\put(160,-15){\makebox(0,0)[t]{$L-1$}} 
\put(167,00){\line(1,0){26}} 
\put(200,00){\circle{15}}
\put(195,-5){\line(1, 1){10}} 
\put(195, 5){\line(1,-1){10}} 
\dottedline{3}(208,0)(232,0) 
\put(240,00){\circle{15}}
\dottedline{3}(250,0)(270,0) 
%
\end{picture}
\end{minipage}
\ee
The Bethe equations are thus  reduced to the simple equation 
\be
\left(\frac{u_{5,k}+i}{u_{5,k}-i}\right)^L = 1,\qquad k = 1, \dots, L-1,
\ee
which is solved by 
\be
u_{5,k} = \cot\frac{\pi\,k}{L}.
\ee
The dualization process can be inverted step by step providing exact polynomials whose roots are the Bethe roots at any finite $L$. 
In particular, one finds for the roots at node 4 and 6 the explicit result~\cite{Rej:2007vm} 
\ba
Q_4(u) &=& \left(u+\frac{i}{2}\right)^L\,\left[\left(u+\frac{3\,i}{2}\right)^L-\left(u-\frac{i}{2}\right)^L\right]
+\left(u-\frac{i}{2}\right)^L\,\left[\left(u-\frac{3\,i}{2}\right)^L-\left(u+\frac{i}{2}\right)^L\right]. \nonumber \\
Q_6(u) &=& \left(u+\frac{3\,i}{2}\right)^L+\left(u-\frac{3\,i}{2}\right)^L-\left(u+\frac{i}{2}\right)^L-\left(u-\frac{i}{2}\right)^L.
\ea
Of course, from the knowledge of $Q_4(u)$ one can prove again the one-loop result Eq.~(\ref{eq:oneloopgamma}). 

Going over to higher orders, we have to work with the long range Bethe equations which are a 
deformation of the one-loop ones. They involve the standard quantities
\be
x(u) = \frac{u}{2}\left(1+\sqrt{1-\frac{2\,g^2}{u^2}}\right),\qquad x^\pm = x\left(u\pm\frac{i}{2}\right),
\ee
and read
\begin{eqnarray}
1&=&\prod_{j=1}^{2L-2}\frac{x_{3,k}-x_{4,j}^{+}}{x_{3,k}-x_{4,j}^{-}},\nonumber\\
\left (\frac{x_{4,k}^{+}}{x_{4,k}^{-}} \right )^{L}&=&\prod_{\substack{j=1 \\j\neq k}}^{2L-2}\frac{x_{4,k}^{+}-x_{4,j}^{-}}{x_{4,k}^{-}-x_{4,j}^{+}}
\frac{1-g^{2}/2\,x_{4,k}^{+}x_{4,j}^{-}}{1-g^{2}/2\,x_{4,k}^{-}x_{4,j}^{+}}\,
\sigma^2(u_{4,k},u_{4,j})
\nonumber\\
&&\times\prod_{j=1}^{2L-3}\frac{x_{4,k}^{-}-x_{3,j}}{x_{4,k}^{+}-x_{3,j}} \prod_{j=1}^{L-1}\frac{x_{4,k}^{-}-x_{5,j}}{x_{4,k}^{+}-x_{5,j}}\prod_{j=1}^{L-3}
\frac{1-g^{2}/2\,x_{4,k}^{-}x_{7,j}}{1-g^{2}/2\,x_{4,k}^{+}x_{7,j}},
\nonumber\\
1&=&\prod_{j=1}^{L-2}\frac{u_{5,k}-u_{6,j}+\frac{i}{2}}{u_{5,k}-u_{6,j}-\frac{i}{2}}\prod_{j=1}^{2L-2}\frac{x_{5,k}-x_{4,j}^{+}}{x_{5,k}-x_{4,j}^{-}}, \\
1&=&\prod_{\substack{j=1\\j \neq k}}^{L-2}\frac{u_{6,k}-u_{6,j}-i}{u_{6,k}-u_{6,j}+i}\prod_{j=1}^{L-1}\frac{u_{6,k}-u_{5,j}+\frac{i}{2}}{u_{6,k}-u_{5,j}-\frac{i}{2}}
\prod_{j=1}^{L-3}\frac{u_{6,k}-u_{7,j}+\frac{i}{2}}{u_{6,k}-u_{7,j}-\frac{i}{2}}, \nonumber\\
1&=&\prod_{j=1}^{L-2}\frac{u_{7,k}-u_{6,j}+\frac{i}{2}}{u_{7,k}-u_{6,j}-\frac{i}{2}}\prod_{j=1}^{2L-2}\frac{1-g^{2}/2\,x_{7,k}x_{4,j}^{+}}{1-g^{2}/2\,x_{7,k}x_{4,j}^{-}},
\nonumber
\end{eqnarray}
where $\sigma^2(u_k,u_j)$ is the dressing phase to be discussed later.

It is possible to perform a {\em partial} dualization of these equations and obtain reduced long-range equations involving roots at nodes 4, 5, 6 only. These are
\begin{eqnarray}
\left(\frac{x^{+}_{4,k}}{x^{-}_{4,k}}\right)^{L}&=&\prod^{2L-2}_{\substack{j=1\\j\neq k}} \frac{x^{-}_{4,k}-x^{+}_{4,j}}{x^{+}_{4,k}-x^{-}_{4,j}}\, 
\frac{1-\frac{g^2}{2\,x^{+}_{4,k}x^{-}_{4,j}}}{1-\frac{g^2}{2\,x^{-}_{4,k} x^{+}_{4,j}}}
\,\sigma^2(u_{4,k},u_{4,j})\, 
\prod^{2L-4}_{j=1} \frac{x^{+}_{4,k}-\widetilde{x}_{5,j}}{x^{-}_{4,k}-\widetilde{x}_{5,j}}\label{all1}, \nonumber\\
1&=&\prod^{L-2}_{j=1}\frac{\widetilde{u}_{5,k}-u_{6,j}+\frac{i}{2}}{\widetilde{u}_{5,k}-u_{6,j}-\frac{i}{2}} \prod^{2L-2}_{j=1}
\frac{\widetilde{x}_{5,k}-x^{+}_{4,j}}{\widetilde{x}_{5,k}-x^{-}_{4,j}}\label{all2},\\
1&=&\prod^{L-2}_{\substack{j=1\\ j \neq k}} \frac{u_{6,k}-u_{6,j}+i}{u_{6,k}-u_{6,j}-i} \prod^{2L-4}_{j=1} 
\frac{u_{6,k}-\widetilde{u}_{5,j}-\frac{i}{2}}{u_{6,k}-\widetilde{u}_{5,j}+\frac{i}{2}}\label{all3}.\nonumber
\end{eqnarray}
Here, $\widetilde{u}_5$ are the $2L-4$ roots dual to $u_5$. At one loop, they are the roots of the polynomial
\begin{eqnarray}
Q_{5}(u)&=&3u^{2L}+(-i+u)^L (-2i+u)^L+(2i+u)^L 
\left((i+u)^L+(-2i+u)^L\right)\nonumber\\
&&-u^L \left((-i+u)^L+(i+u)^L+2(-2i+u)^L+2(2i+u)^L\right).
\end{eqnarray}
The roots at nodes 4 and 6 are still given at one-loop by the previous polynomials.  At generic $g$, the anomalous dimension 
is obtained from the roots $u_{4,k}(g)$ alone and reads
\be
\gamma_L(g) = 2\,L + g^2\,\sum_{k=1}^{K_4}\left(\frac{i}{x^+(u_{4,k})}-\frac{i}{x^-(u_{4,k})}\right).
\ee
Finally, let us consider the dressing phase. It enters the calculation starting from four loops. 
Its general form is discussed in~\cite{Beisert:2006ez}. The terms relevant for a computation up to five loops
are simply 
\be
\sigma^2(u, u') = e^{i\,\vartheta(u, u')},
\ee
where 
\ba
\vartheta(u,u') &=& (\zeta_3\,g^6-5\,\zeta_5\,g^8)\,(q_2(u)\,q_3(u')-q_2(u')\,q_3(u))+\cdots, \\ \nonumber \\
q_2(u) &=& i\,\left(\frac{1}{x^+(u)}-\frac{1}{x^-(u)}\right),  \quad q_3(u) = \frac{i}{2}\,\left(\frac{1}{x^+(u)^2}-\frac{1}{x^-(u)^2}\right). \nonumber
\ea
The coefficients $\zeta_n$ are transcendental sums $\zeta_n = \sum_{\ell=1}^\infty \ell^{-n}$.

\subsection{The one-loop Bethe roots: some numerics at large but finite $L$}

The one-loop Bethe roots are the zeroes of the polynomials $Q_{4,5,6}(u)$. It is instructive to study them at large $L$ comparing with the 
results of~\cite{Rej:2007vm} obtained in the $L\to\infty$ limit. First the (dual) roots $u_5$. They are complex. We show them at $L=100, 200, 350, 500$
in Fig.~(\ref{fig:u5}). As predicted, most of them are distributed along two segments with $\mbox{Im}\, u_5 = \pm\frac{i}{2}$. 
Apart from these roots, other ones are scattered in the complex plane according to a nice regular pattern. 
To understand these roots, we look for a Bethe root admitting the expansion 
\be
u = L\,\left(x_0 + \frac{x_1}{L^{1/2}} + \frac{x_2}{L} + \cdots\right).
\ee
A roots with leading behavior $u\sim L$ is called {\em extremal} in~\cite{Rej:2007vm}. Replacing this expansion in $Q_6(u)$ , we obtain a well defined large $L$ 
expansion for the ratio
\be
R(x_0, x_1, \dots; L) = \frac{Q_6(u)}{u^{2L}}.
\ee
The leading term is 
\be
R = 64\,\cos^2\frac{1}{2\,x_0}\,\sin^4\frac{1}{2\,x_0} + {\cal O}(L^{-1/2}),
\ee
leading to $x_0 = \frac{1}{n\,\pi}$ for any integer $n$. Considering separately the cases $n$ even/odd and expanding at higher order in $L^{-1/2}$
one finds the solutions (in the first quadrant)
\ba
\alpha_{\pm, k} &=& \frac{1}{(2\,k+1)\,\pi}\left(L\pm\sqrt\frac{L}{2}\right)+{\cal O}\left(L^{-1/2}\right),\qquad k = 0, 1, 2, \dots, \\
\beta_{\pm, k} &=& \frac{1}{2\,k\,\pi}\left(L\pm\frac{1}{2}\sqrt{L}\,\sqrt{3\pm\,i\,\sqrt{15}}\pm\frac{i}{\sqrt{15}}\right)+{\cal O}\left(L^{-1/2}\right),
\qquad k = 1, 2, \dots.
\ea
The other roots are related by reflection with respect to the coordinate axis. For large $L$, the $\alpha$-roots appear in real close pairs. These pairs are closer to the origin 
as $k$ is increased. In general, for a given $L$, only a finite number of such pairs is well approximated by the above formula. The $\beta$-roots have an imaginary part
and also appear in close pairs. In Fig.~(\ref{fig:u5}) we draw crosses at the first $\alpha$ and $\beta$ pairs.

The roots $u_{4,n}$ and $u_{6,n}$ are real. Their density is defined in the $L\to\infty$ continuum limit as
$\rho(u) = dn/du$ and the analytical prediction is 
\be
\rho_4(u) = \frac{1}{2\,\pi}\left(\frac{1}{u^2+\frac{1}{4}}+\frac{3}{u^2+\frac{9}{4}}\right),\qquad 
\rho_6(u) = \frac{1}{2\,\pi}\frac{3}{u^2+\frac{9}{4}}.
\ee
In the discrete case at finite $L$, we can plot the points
\be
\left(\frac{u_n+u_{n+1}}{2}, \frac{4\,L}{u_{n+1}-u_n}\right).
\ee
The result is shown in Figs.~(\ref{fig:u4},\ref{fig:u6}) for the Bethe roots at $L=200$. The agreement is quite good in the case
of $u_4$. For $u_6$, we observe a deviation in the tails of the distribution at large $|u_6|$. It can be understood as in the above discussion of extremal
$u_5$ roots.

\section{Perturbative expansion of the long-range Bethe equations}

Starting from the exact ({\em i.e.} known with arbitrarily high precision) one-loop Bethe roots we can 
make a perturbative expansion in even powers of $g$
\be
u_{a, k} = \sum_{n=0}^\infty g^{2\,n}\,u_{a,k}^{(n)},\qquad a = 4, 5, 6,\qquad k = 1, \dots, K_a,
\ee
where we relabel $\widetilde{u}_5 \equiv u_5$, $\widetilde{K}_5 \equiv K_5$. The explicit five loop 
expansion of the anomalous dimension can be compared with the results of~\cite{Rej:2007vm} up to $L=8$.
We have extended the calculation up to $L=60$. The results at five loops for $L\le 20$ are shown in Appendix~(A).

The expansion is a rational combination of $1$, $\zeta_3$ and $\zeta_5$. As we mentioned, the zero and one loop 
results are proportional to $L$. In general, it is convenient to redefine
\be
\gamma_L(g) = L\, \left(2 + 3\,g^2 + \sum_{n\ge 2} c_n(L)\, g^{2\,n}\right).
\ee
We now show that it is possible to provide simple closed expressions for the non-trivial functions $c_n(L)$.
As a constraint, we must meet the exact expansion in the $L\to\infty$ limit obtained in~\cite{Rej:2007vm} and reading at five loops
\ba
c_2(\infty) &=& -\frac{51}{8},\nonumber\\
c_3(\infty) &=& \frac{393}{16},\\
c_4(\infty) &=& -\frac{59487}{512}-\frac{27}{4}\,\zeta_3, \nonumber\\
c_5(\infty) &=& \frac{632661}{1024}+\frac{1665}{32}\,\zeta_3 + \frac{135}{4}\,\zeta_5.\nonumber
\ea
As a general remark, it is instructive to plot  the numerical values of $c_n(L)$ at the first values of $L$. 
Indeed, it is immediately clear that factors $(-1)^L$ can appear in the closed formula for $c_n(L)$.
Therefore, we shall analyze the odd and even $L$ cases separately.

\subsection{Two loops}

For odd $L = 5, 7, 9, \dots$, we subtract the asymptotic value $c_2(\infty)$ and rescale to find
\be
\frac{8}{3}(c_2(L) - c_2(\infty)) = 
-\frac{1}{5},-\frac{1}{21},-\frac{1}{85},-\frac{1}{341},-\frac{1}{1365},-\frac{1}{5461},-\frac{1}{21845},-\frac{1}{87381},-\frac{1}{349525}, \dots
\ee
A careful inspection reveals that the denominators are simply related to powers of 2 minus one. The precise formula is easily found
and reads
\be
c_2(L) = -\frac{51}{8}-\frac{9}{8}\,\frac{1}{2^{L-1}-1},\qquad L\ \mbox{odd}.
\ee
We checked it for all the $L$ that we have explored. Remarkably, it works also for the even $L$ case
if the sign of the term $\sim 2^L$ is changed. The final formula is thus
\be
\label{eq:twoloop}
c_2(L) = -\frac{51}{8}+\frac{9}{8}\,\frac{1}{(-1)^L\,2^{L-1}+1}.
\ee
This simple result is rather remarkable. It holds at finite $L$ and predict exponentially suppressed deviations from the 
trivial linear scaling of the anomalous dimension $\gamma\sim L$, valid up to the one-loop level. Is it possible to obtain a similar
result for the next three loop contribution ? 

\subsection{Three loops}

Following the strategy adopted in the two-loop case, we start again from odd $L = 5, 7, 9, \dots$ and evaluate 
\ba
\lefteqn{c_3(L)-c_3(\infty) = 
+\frac{111}{1600},-\frac{425}{3136},-\frac{3628101}{39304000},} && \\
&& \qquad\qquad\qquad\qquad -\frac{9904623}{230701504},-\frac{7874523}{463736000},-\frac{63804855621}{10423090379584}, \dots .\nonumber
\ea
This sequence appears to be definitely non trivial and much more complicated than the two-loop case. In particular, the signs are not 
definite and the denominators do not have simple factorization properties. However, the sequence enjoys a remarkable property. 
If we multiply it by $(2^{L-1}-1)^3$ and apply a constant scaling,
we find 
\be
\frac{2^6}{3^4}\,(2^{L-1}-1)^3\, (c_3(L)-c_3(\infty)) = 
185, -26775, -1209367, -36316951, -921319191, 
\ee
$$
\ \ -21268285207, -461958727447, -9613145655063, -193758643734295,\dots.
$$
Indeed, the sequence is integer for all considered $L$. As a second feature, one can plot the following function of $L$
\be
(2^{L-1}-1)\, (c_3(L)-c_3(\infty)),
\ee
and it turns out to be curve quite close to a quadratic parabola. From these two features, it is natural to look for 
a closed formula of the form 
\be
c_3(L)-c_3(\infty) = \frac{1}{(2^L-2)^3}\sum_{p=0}^2\, 2^{p\, L}\, \sum_{q=0}^2 c_{p,q}\,L^q.
\ee
Indeed, it turns out that all the three loop results at odd $L$ are reproduced by 
\be
c_3(L) = \frac{393}{16} + \frac{-9\cdot 2^{2 L} \left(9\, L^2-33\, L-104\right)-18\cdot 2^{L} \left(9\, L^2+15\, L+202\right)+3528}{64 \left(-2+2^L\right)^3},\ L\ \mbox{odd}.
\ee
Looking back at Eq.~(\ref{eq:twoloop}), there is a striking similarity suggesting an all order structure. In particular, 
the same formula works for even $L$, if we apply the modification rules 
\be
2^{2\,p\,L}\to 2^{2\,p\,L},\qquad 2^{(2\,p+1)\,L}\to -(-1)^L\,  2^{(2\,p+1)\,L}.
\ee
The general formula is then 
\be
c_3(L) = \frac{393}{16} + \frac{9\cdot 2^{2 L} \left(9\, L^2-33\, L-104\right)-18\cdot(-1)^L\, 2^{L} \left(9\, L^2+15\, L+202\right)-3528}
{64\cdot 8\, \left[(-1)^L\,2^{L-1}+1)\right]^3}.
\ee

\subsection{Four loops}

At four loops, we attempt to repeat the game. The only new feature is the transcendental contribution from the dressing phase.
This is a piece of $c_4(L)$ proportional to $\zeta_3$. From the numerics, it is independent on $L$ and reads
\be
c_4(L) = c_4^{(0)}(L) + c_4^{(3)}(L)\,\zeta_3,\qquad c_4^{(3)}(L)\equiv  -\frac{27}{4}.
\ee
The $c_4^{(0)}(L)$ is a rational contribution with properties quite analogous to those of $c_3(L)$. In particular, for odd $L$
\ba
(i)\ && \frac{2^8}{3^7}\,(2^{L-1}-1)^5\, (c_4^{(0)}(L)-c_4^{(0)}(\infty)) \in \mathbb{N}, \\
(ii)\ && (2^{L-1}-1)\, (c_4^{(0)}(L)-c_4^{(0)}(\infty)) \sim L^4,\ \ \mbox{for $L\to\infty$}.
\ea
Again, it is natural to postulate from (i) and (ii) the closed formula
\be
c_4^{(0)}(L)-c_4^{(0)}(\infty) = \frac{1}{(2^L-2)^5}\sum_{p=0}^4\, 2^{p\, L}\, \sum_{q=0}^4 d_{p,q}\,L^q.
\ee
Replacing the explicit anomalous dimensions in this formula we find that indeed it is satisfied by all considered (odd) $L$ with 
coefficients $d_{p,q}$ giving 
\be
c_4^{(0)}(L) = -\frac{59487}{512} + \frac{2^{4 L}\,Q_{4,4} + 2^{3 L}\, Q_{4,3} + 2^{2 L}\, Q_{4,2} + 2^{L}\, Q_{4,1} -1335168}{2^{15}\left(2^{L-1}-1\right)^5}
,\quad L \ \mbox{odd},
\ee
where the $Q$ polynomials are 
\ba
Q_{4,1} &=& -72 \left(27\, L^4+90\, L^3-1485\, L^2-2004\, L-38456\right), \nonumber\\
Q_{4,2} &=& -108 \left(99\, L^4-18 \, L^3+513\, L^2+2958\, L+19924\right), \\
Q_{4,3} &=& -18 \left(297\, L^4-1080\, L^3+1647\, L^2-12060\, L-41264\right), \nonumber\\
Q_{4,4} &=& -9\,\left(27\, L^4-252\, L^3-1053\, L^2+5190\, L+10676\right).\nonumber
\ea
The case $L$ even is obtained changing the sign of $2^L$ in the powers $(2^L)^p$ and correcting with a shift in the boundary case $L=4$. 
The final result is 
\ba
c_4^{(0)}(L) &=& -\frac{5}{64}\,\delta_{L,4} -\frac{59487}{512} + \nonumber\\
&& - \frac{2^{4 L}\,Q_{4,4} -(-1)^L\, 2^{3 L}\, Q_{4,3} + 2^{2 L}\, Q_{4,2} -(-1)^L\, 2^{L}\, Q_{4,1} -1335168}{2^{15}\left[(-1)^L\,2^{L-1}+1\right]^5}.
\ea
The above shift, as well as the corrections appearing in the five loop formula
(\ref{five}) below, are possibly related to short wrapping effects - the lack
of the asymptotic conditions prevents in the boundary cases the validity of the
Bethe equations.

\subsection{Five loops}

At five loops, we have a more complicated dressing contribution with two different transcendentality terms 
\be
c_5(L) = c_5^{(0)}(L )+ c_5^{(3)}(L)\,\zeta_3+ c_5^{(5)}(L)\,\zeta_5,\qquad c_5^{(0, 3, 5)}(L)\in\mathbb{Q}.
\ee
The maximum transcendentality $c_5^{(5)}$ is independent on $L$
\be
c_5^{(5)}(L) = \frac{135}{4}.
\ee
Repeating the above heuristic analysis for the other terms we find for the transcendentality 3 term
\be
c_5^{(3)}(L) = \frac{1665}{32}-\frac{3}{8}\,\delta_{L,4} + \frac{81\cdot(-1)^L\,2^L\,(L-4)-648}{2^7\left[(-1)^L\,2^{L-1}+1\right]^2}. 
\ee
The purely rational term has the representation
\be
c_5^{(0)}(L)-c_5^{(0)}(\infty) = \frac{1}{(2^L-2)^7}\sum_{p=0}^6\, 2^{p\, L}\, \sum_{q=0}^6 e_{p,q}\,L^q,
\ee
with the explicit final result, holding for even or odd $L$
\ba
c_5^{(0)} &=& \frac{632661}{1024}+\frac{14987}{12288}\,\delta_{L,4}-\frac{333}{4096}\,\delta_{L,5} + \frac{G(L)}{2^{22}\,\left[(-1)^L\,2^{L-1}+1\right]^7}, \\
G(L) &=& \sum_{p=0,6} (-1)^{p\,(L+1)}\, 2^{p\,L}\,Q_{5,p},
\label{five}
\ea
where
\ba
Q_{5,0} &=& -2^{11}\cdot 3^2\, (432\, L+56639), \\
Q_{5,1} &=& 2^5\cdot 3^4\, \left(9 \,L^6+45 \,L^5-1581 \,L^4-4113 \,L^3+39492 \,L^2+53316 \,L+1253696\right), \nonumber\\
Q_{5,2} &=& 2^4\cdot 3^3\, \left(1539 \,L^6+1755 \,L^5-37503 \,L^4+41409 \,L^3-370980 \,L^2-961116 \,L-9751792\right), \nonumber\\
Q_{5,3} &=& 2^4\cdot 3^2\,\left(12231 \,L^6-17172 \,L^5+68067 \,L^4+158976 \,L^3+358722 \,L^2+3589416 \,L+20219128\right), \nonumber\\
Q_{5,4} &=& 2^3\cdot 3^5\,\left(453 \,L^6-1956 \,L^5+3769 \,L^4-15096 \,L^3+13278 \,L^2-163616 \,L-582008\right), \nonumber\\
Q_{5,5} &=& 2\cdot 3^3\,\left(1539 \,L^6-13095 \,L^5-10611 \,L^4+82683 \,L^3-290952 \,L^2+1783716 \,L+4340656\right), \nonumber\\
Q_{5,6} &=& 9 \left(81 \,L^6-1377 \,L^5-6129 \,L^4+103653 \,L^3+195912 \,L^2-1277388 \,L-2247232\right).\nonumber 
\ea
The extension to higher loops seems to be a computational issue. One has to generate a large enough number of 
terms in $c_n(L)$ and must check that an Ansatz similar to the previous ones matches it. 

\section{Large $L$ expansion of $\gamma_L(g)$}

The five-loop results described in the previous sections are valid at finite $L$. Nevertheless, it is interesting to 
look at the dominant terms at large $L$. As remarked in the Introduction, the resulting expression can admit a thermodynamical interpretation.
Collecting the formulae for $c_n$ and expanding at large $L$, we find 
\be
\frac{\gamma_L(g)}{L} = f_0(g) + f_1(g, L)\, e^{-L\,\log\,2} + f_2(g, L)\, e^{-2\,L\,\log\,2}+ f_3(g, L)\, e^{-3\,L\,\log\,2} + \cdots .
\ee
The leading term agrees by construction with the result of ~\cite{Rej:2007vm}
\be
f_0(g) = 2+ 3\,g^2-\frac{51}{8}\,g^4 + \frac{393}{16}\,g^6 + \left(-\frac{27 \zeta_3 }{4}-\frac{59487}{512}\right)\,g^8 + 
\left(\frac{1665 \zeta_3 }{32}+\frac{135 \zeta_5 }{4}+\frac{632661}{1024}\right)\,g^{10} + \dots .
\ee
The first exponentially suppressed term has a prefactor
\ba
f_1(g, L) &=& -\frac{9}{4}\,g^4 + \left(-\frac{81 L^2}{64}+\frac{297 L}{64}+\frac{117}{8}\right)\,g^6 + \nonumber \\
&& + \left(-\frac{243 L^4}{1024}+\frac{567
   L^3}{256}+\frac{9477 L^2}{1024}-\frac{23355 L}{512}-\frac{24021}{256}\right)\,g^8 + \nonumber\\
&& + \left(-\frac{729 L^6}{32768}+\frac{12393 L^5}{32768}+\frac{55161
   L^4}{32768}-\frac{932877 L^3}{32768}-\frac{220401 L^2}{4096}+ \right. \nonumber\\
&& \left. + \left(\frac{2874123}{8192}-\frac{81 \zeta_3 }{32}\right) L+\frac{81 \zeta_3
   }{8}+\frac{316017}{512}\right)\,g^{10} + \dots .
\ea
At order ${\cal O}(g^{2\,n})$, the leading power of the length is $L^{2\,n-4}$ and comes always in transcendentality 0 terms unrelated to dressing.
The large $L$ limit of $f_1(g, L)$ can be compactly written as 
\be
f_1(g, L) = -\frac{9}{4}\,g^4\left(1+z^2+\frac{z^4}{3}+\frac{z^6}{18}+\cdots\right),\qquad z = \frac{3}{4}\,L\,g,
\ee
 and in particular, given the absence of transcendental contributions, do not depend on the dressing phase. It seems reasonable that this structural properties could persist 
at all orders.

\section{Discussion and Conclusions}
\label{sec:conc}

In this paper, we have considered the chiral operator $\mbox{Tr}\,{\cal F}^L$ in ${\cal N}=4$ SYM.
At one-loop, it scales with a definite anomalous dimension $\gamma_L$ proportional to $L$. At two-loops and beyond, it mixes
with the other $\mathfrak{psu}(2,2|4)$ fields. The length $L$ is no more a conserved quantity 
and $\gamma_L/L$ is not constant. In principle, this ratio is not expected to be expressed by a simple expression at finite $L$. 
One would just resort to compute systematically its corrections at large $L$.

Nevertheless, the main result of this paper shows that some unexpected structure exists at finite $L$.
We have been able to provide a closed form for $\gamma_L/L$ up to five-loops.
Radiative corrections follow a simple pattern order by order in perturbation theory, including transcendental dressing effects.
They are sensitive to the parity of $L$ and are exponentially suppressed as $L\to\infty$.

A closed formula for the multi-loop size dependence is a remarkable fact 
that has no counterpart in existing calculations for other operators in the various subsectors
of ${\cal N}=4$ SYM. It can be due to the simplicity of the considered operator or could hint to 
some hidden relation obeyed by the anomalous dimensions as a function of $L$. The closed formulae 
are a mere conjecture, although with a strong empirical basis. It is clear that a (dis)proof
would be certainly enlightening.

In the large volume regime our results read
\be
\label{eq:largevolume}
\frac{\gamma_L(g)}{L} = f_0(g) + g^4 h(g\,L)\, e^{-L\,\log\,2} + \cdots.
\ee
Eq.~(\ref{eq:largevolume}) claims that starting at two-loops, exponentially suppressed corrections appear
with a $g$ independent correlation length $\xi = 1/\log\,2$ and the combination $g\,L$ as
a natural scaling variable for the prefactor.
It would be interesting to understand such features from the point of view
of the spin-chain interpretation of the dilatation operator $H$. We emphasize that the ${\cal O}(2^{-L})$
corrections have nothing to do with much smaller ${\cal O}(\lambda^L)$ wrapping effects. 
A natural explanation for the exponential corrections could take into account length-changing processes
as suggested in~\cite{Rej:2007vm}. An explicit two-loop calculation of $H$ would be important to clarify
these issues.

We conclude with a remark concerning the dressing phase $\vartheta$. Currently, this is a well understood ingredient appearing in the 
$S$-matrix. However, it would be very nice to classify the special kind of interactions that are associated with it in the dilatation 
operator. A relevant step in this direction has been recently described in~\cite{Beisert:2007hz} where it is linked to 
so-called maximal reshuffling interactions. In our investigation, the special feature of dressing effects is that they are
subleading at large $L$ and up to five-loops. Transcendental contributions drop out from the function $f_1(L, g)$ 
being characterized by subdominant powers of the length $L$.

\acknowledgments
We thank M.~Staudacher for suggestions and comments. We also thank 
G.~F.~De Angelis for discussions.

\vfill

\appendix
\section{Five loop anomalous dimensions for $L\le 20$}
\ba
\gamma_{4}(g) &=& 8+12\,g^2-25\,g^4+\frac{1515}{16}\,g^6+\left(-\frac{513937}{1152}-27\,\zeta_3 \right)\,g^8\nonumber\\
&& +
\left(\frac{22129823}{9216}+\frac{1651}{8}\,\zeta_3 + 135\,\zeta_5 \right)\,g^{10} +\cdots,\nonumber \\ 
\gamma_{5}(g) &=& 10+15\,g^2-\frac{129}{4}\,g^4+\frac{39411}{320}\,g^6+\left(-\frac{7346253}{12800}-\frac{135}{4}\,\zeta_3 \right)\,g^8 \nonumber\\
&& +\left(\frac{1539949881}{512000}+\frac{8307}{32}\,\zeta_3 + \frac{675}{4}\,\zeta_5 \right)\,g^{10} +\cdots,\nonumber \\ 
\gamma_{6}(g) &=& 12+18\,g^2-\frac{837}{22}\,g^4+\frac{6278355}{42592}\,g^6+\left(-\frac{29266837713}{41229056}-\frac{81}{2}\,\zeta_3 \right)\,g^8 \nonumber\\
&& +\left(\frac{76857234976107}{19954863104}+\frac{605205}{1936}\,\zeta_3 + \frac{405}{2}\,\zeta_5 \right)\,g^{10} +\cdots,\nonumber \\ 
\gamma_{7}(g) &=& 14+21\,g^2-\frac{179}{4}\,g^4+\frac{76603}{448}\,g^6+\left(-\frac{181131695}{225792}-\frac{189}{4}\,\zeta_3 \right)\,g^8 \nonumber\\
&& +\left(\frac{8959397257}{2107392}+\frac{11641}{32}\,\zeta_3 + \frac{945}{4}\,\zeta_5 \right)\,g^{10} +\cdots,\nonumber \\ 
\gamma_{8}(g) &=& 16+24\,g^2-\frac{2190}{43}\,g^4+\frac{125533809}{636056}\,g^6+\left(-\frac{8840715968859}{9408540352}-54\,\zeta_3 \right)\,g^8\nonumber \\
&& +\left(\frac{346753221469919673}{69585564443392}+\frac{3080871}{7396}\,\zeta_3 + 270\,\zeta_5 \right)\,g^{10} + \cdots,\nonumber \\ 
\gamma_{9}(g) &=& 18+27\,g^2-\frac{19521}{340}\,g^4+\frac{8655987591}{39304000}\,g^6+\left(-\frac{2364798793587021}{2271771200000}-\frac{243}{4}\,\zeta_3 \right)\,g^8\nonumber\\
&& +\left(\frac{73033337654861466627}{13130837536000000}+\frac{108214623}{231200}\,\zeta_3 + \frac{1215}{4}\,\zeta_5 \right)\,g^{10} + \cdots,\nonumber \\ 
\gamma_{10}(g) &=& 20+30\,g^2-\frac{7265}{114}\,g^4+\frac{486455845}{1975392}\,g^6+\left(-\frac{179336215108445}{154033166592}-\frac{135}{2}\,\zeta_3 \right)\,g^8\nonumber\\
&& +\left(\frac{4103422374381475165}{667271677676544}+\frac{27055595}{51984}\,\zeta_3 + \frac{675}{2}\,\zeta_5 \right)\,g^{10} + \cdots,\nonumber \\ 
\gamma_{11}(g) &=& 22+33\,g^2-\frac{8697}{124}\,g^4+\frac{5656701069}{20972864}\,g^6+\left(-\frac{206094402320199}{161239378432}-\frac{297}{4}\,\zeta_3 \right)\,g^8\nonumber\\
&& +\left(\frac{512022650767888797357}{74996304653805568}+\frac{17597781}{30752}\,\zeta_3 + \frac{1485}{4}\,\zeta_5 \right)\,g^{10} + \cdots,\nonumber \\ 
\gamma_{12}(g) &=& 24+36\,g^2-\frac{52245}{683}\,g^4+\frac{1504241745363}{5097791792}\,g^6+\left(-\frac{26503161491873431953}{19024510362066304}-81\,\zeta_3 \right)\,g^8\nonumber\\
&& +\left(\frac{262052003573439955673753835}{35498899257159792346624}+\frac{2330333685}{3731912}\,\zeta_3 + 405\,\zeta_5 \right)\,g^{10} + \cdots,\nonumber \\ 
\gamma_{13}(g) &=& 26+39\,g^2-\frac{11603}{140}\,g^4+\frac{11382640977}{35672000}\,g^6+\left(-\frac{61846229508401447}{40901515200000}-\frac{351}{4}\,\zeta_3 \right)\,g^8\nonumber\\
&& +\left(\frac{4197004793411747026501}{521085303648000000}+\frac{26513707}{39200}\,\zeta_3 + \frac{1755}{4}\,\zeta_5 \right)\,g^{10} + \cdots,\nonumber \\ 
\gamma_{14}(g) &=& 28+42\,g^2-\frac{487473}{5462}\,g^4+\frac{224231872961943}{651801084512}\,g^6\nonumber\\
&& +\left(-\frac{63195216734569435123965}{38890942307856038656}-\frac{189}{2}\,\zeta_3 \right)\,g^8 \nonumber\\
&& +\left(\frac{10018082258604066826691362529007}{1160250749448653889305611264}+\frac{86929777809}{119333776}\,\zeta_3 + \frac{945}{2}\,\zeta_5 \right)\,g^{10} + \cdots,\nonumber 
\ea
\ba 
\gamma_{15}(g) &=& 30+45\,g^2-\frac{2088855}{21844}\,g^4+\frac{3839300288893665}{10423090379584}\,g^6\nonumber\\
&& +\left(-\frac{4337543080186955113152555}{2486742653840334490112}-\frac{405}{4}\,\zeta_3 \right)\,g^8\nonumber\\
&& +\left(\frac{2751363092765611488400885340297145}{296643740062996423913557649408}+\frac{744805296585}{954320672}\,\zeta_3 + \frac{2025}{4}\,\zeta_5 \right)\,g^{10} + \cdots,\nonumber \\ 
\gamma_{16}(g) &=& 32+48\,g^2-\frac{371380}{3641}\,g^4+\frac{75888854970083}{193073214884}\,g^6\nonumber\\
&& +\left(-\frac{342362914175507036271937}{184287501648332409888}-108\,\zeta_3 \right)\,g^8\nonumber\\
&& +\left(\frac{10731015336070384155651542820939}{1085812212950776269479306368}+\frac{22072903471}{26513762}\,\zeta_3 + 540\,\zeta_5 \right)\,g^{10} + \cdots,\nonumber \\ 
\gamma_{17}(g) &=& 34+51\,g^2-\frac{557049}{5140}\,g^4+\frac{963880152452127}{2308544648000}\,g^6\nonumber\\
&& +\left(-\frac{1024443318894708985760901}{518422022549556800000}-\frac{459}{4}\,\zeta_3 \right)\,g^8\nonumber\\
&& +\left(\frac{122265300022822344725447154392091}{11642027096907730208288000000}+\frac{46737698103}{52839200}\,\zeta_3 + \frac{2295}{4}\,\zeta_5 \right)\,g^{10} + \cdots,\nonumber \\ 
\gamma_{18}(g) &=& 36+54\,g^2-\frac{10027071}{87382}\,g^4+\frac{1180027266090140025}{2668860863627872}\,g^6\nonumber\\
&& +\left(-\frac{85208553177354546108032208051}{40756782343071289195379456}-\frac{243}{2}\,\zeta_3 \right)\,g^8\nonumber\\
&& +\left(\frac{3461778112225195512629525977023541804641}{311203054756192480704870130697145344}+\frac{28604992980447}{30542455696}\,\zeta_3 \right. \nonumber\\
&& \left. + \frac{1215}{2}\,\zeta_5 \right)\,g^{10} 
+ \cdots,\nonumber \\ 
\gamma_{19}(g) &=& 38+57\,g^2-\frac{742739}{6132}\,g^4+\frac{681479838093317}{1460288902464}\,g^6\nonumber\\
&& +\left(-\frac{3455324577868750102309487}{1564904852245239998976}-\frac{513}{4}\,\zeta_3 \right)\,g^8\nonumber\\
&& +\left(\frac{1093345538309226287514039899518949}{93167530542473483540155557888}+\frac{74344962037}{75202848}\,\zeta_3 + \frac{2565}{4}\,\zeta_5 \right)\,g^{10} + \cdots,\nonumber \\ 
\gamma_{20}(g) &=& 40+60\,g^2-\frac{22282275}{174763}\,g^4+\frac{41954422215573649845}{85402081606607152}\,g^6\nonumber\\
&& +\left(-\frac{48481600634126052932187928286535}{20866875547860781826069765504}-135\,\zeta_3 \right)\,g^8\nonumber\\
&& +\left(\frac{31509537567806883433426191930418295740388445}{2549273313592296157452673880096407176704}+\frac{254263222600515}{244336849352}\,\zeta_3 \right. \nonumber\\
&& \left. + 675\,\zeta_5 \right)\,g^{10} 
+ \cdots,\nonumber \\ 
\ea

\newpage

\vskip 2cm
\FIGURE{\epsfig{file=u5.eps, width=16cm}
        \bigskip\bigskip
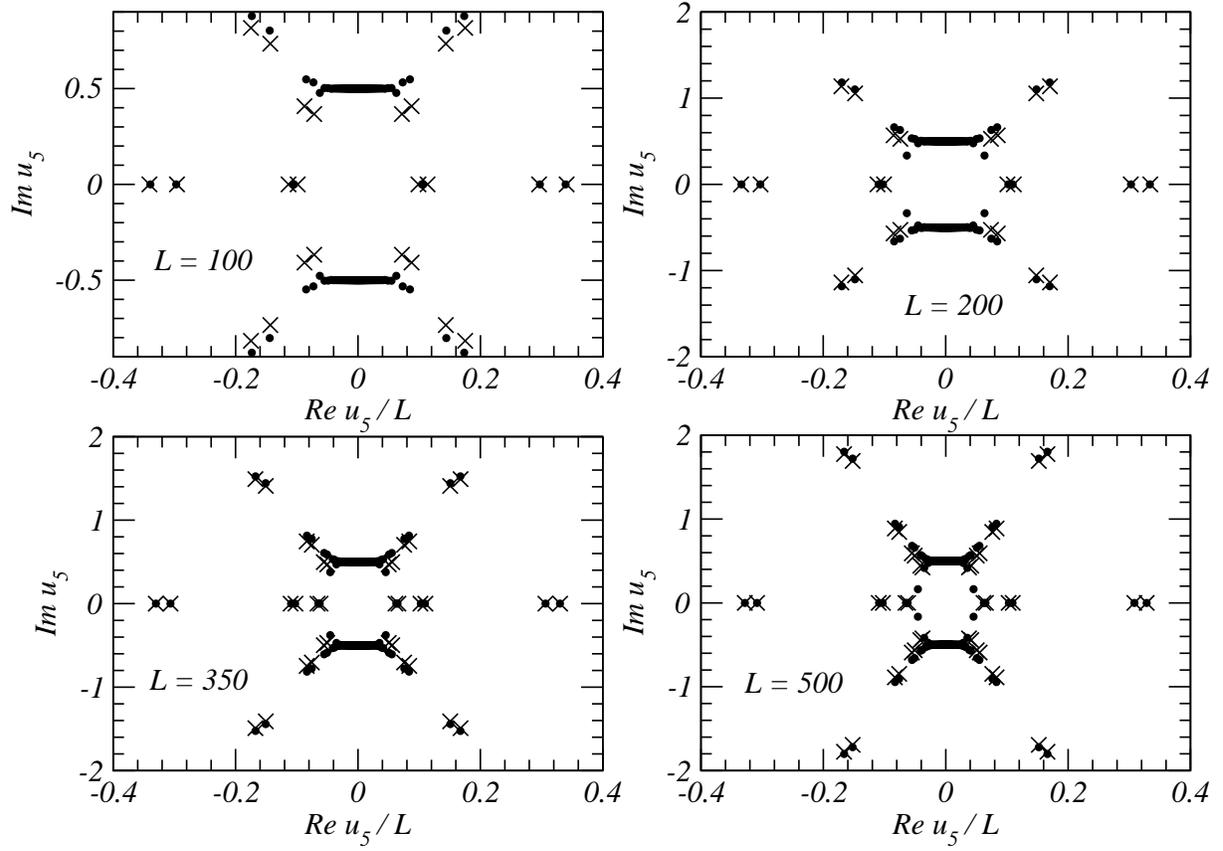
        \caption{Dual Bethe roots $u_5$ computed at one-loop with  $L=100, 200, 350, 500$.
         Crosses on the $x$ axis are pairs of $\alpha_\pm$ extremal roots. Crosses with non zero imaginary part are $\beta_\pm$ roots.}
        \label{fig:u5}}

\newpage
\vskip 2cm
\FIGURE{\epsfig{file=u4.density.eps, width=15cm}
        \bigskip\bigskip
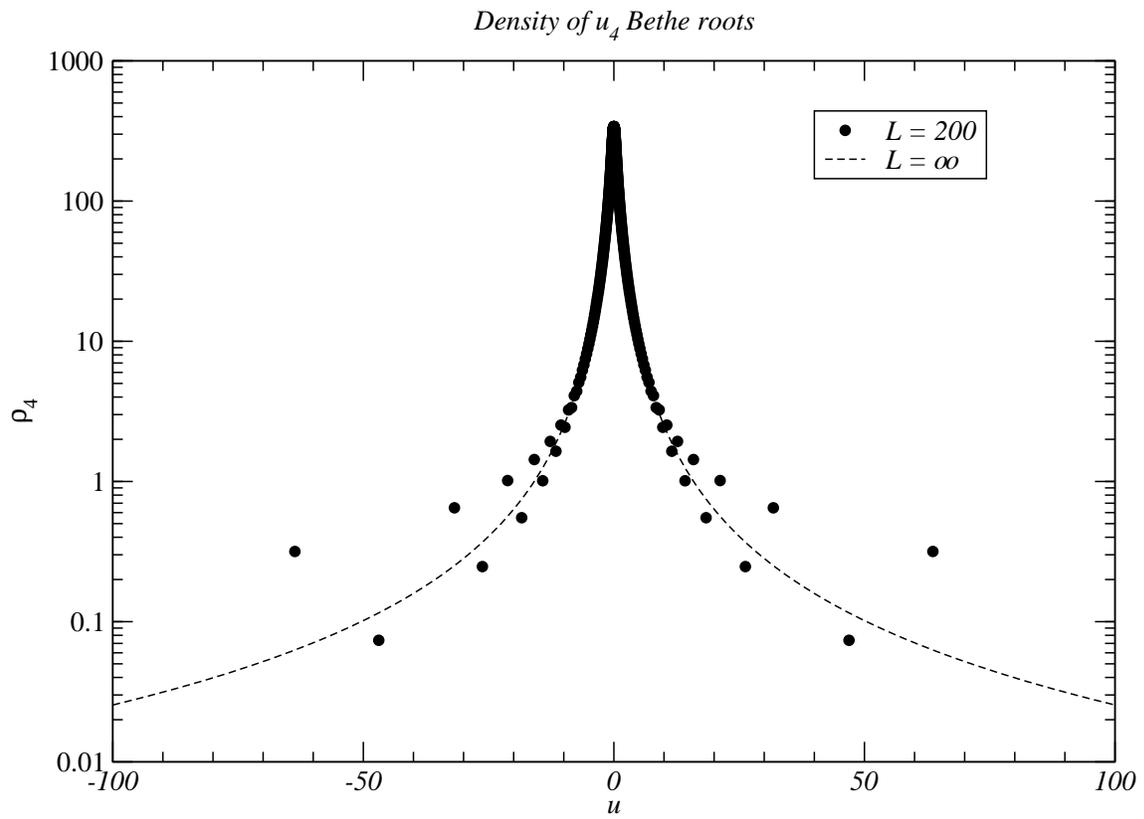
        \caption{Density of Bethe roots $u_4$ from the analytical prediction $\rho_4$ and from the numerical roots at $L=200$.}
        \label{fig:u4}}

\newpage
\vskip 2cm
\FIGURE{\epsfig{file=u6.density.eps, width=15cm}
        \bigskip\bigskip
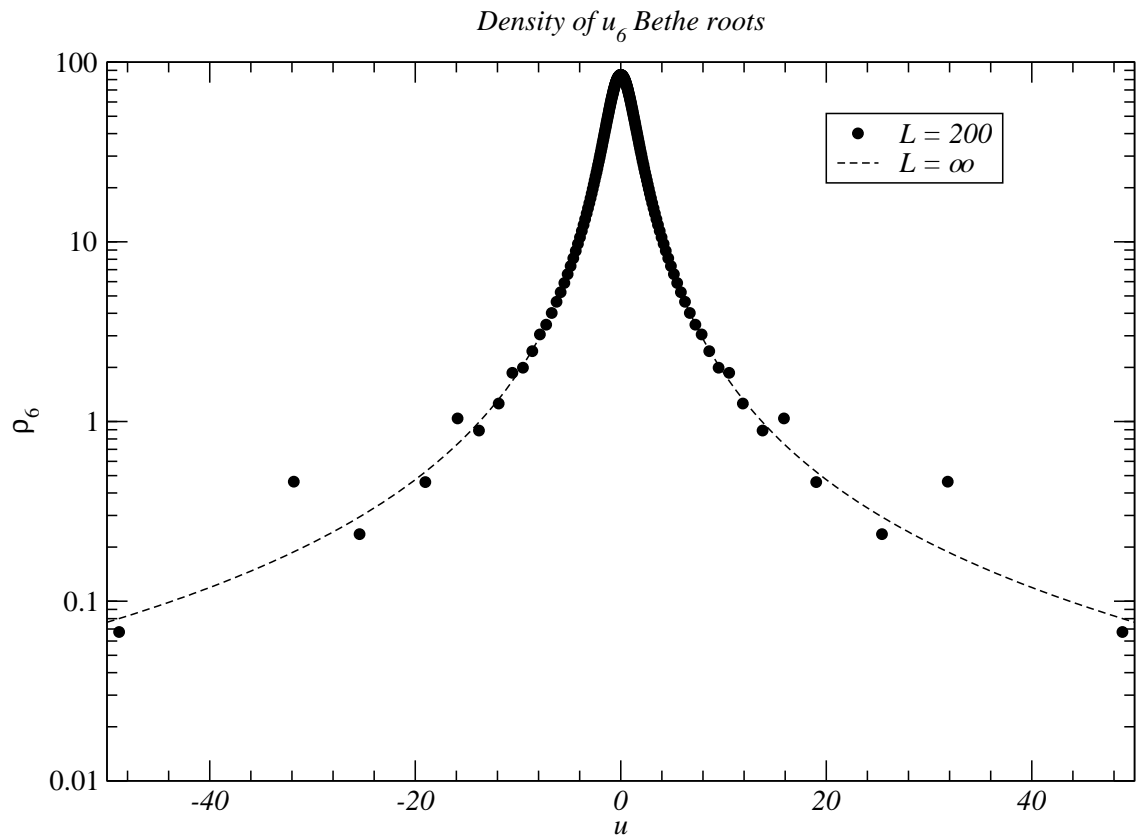
        \caption{Density of Bethe roots $u_6$ from the analytical prediction $\rho_6$ and from the numerical 
        roots at $L=200$.}
        \label{fig:u6}}


\begin{thebibliography}{99}

\bibitem{Belitsky:2004cz}
  A.~V.~Belitsky, V.~M.~Braun, A.~S.~Gorsky and G.~P.~Korchemsky,
  {\em Integrability in QCD and beyond},
  To be published in the memorial volume {\em From Fields to Strings: Circumnavigating Theoretical Physics}, World Scientific, 2004. 
  Dedicated to the memory of Ian Kogan. 
  Int.\ J.\ Mod.\ Phys.\  A {\bf 19}, 4715 (2004)
  [arXiv:hep-th/0407232].

\bibitem{QCD-int}
  L.~N.~Lipatov,
  {\em High-energy asymptotics of multicolor QCD and exactly solvable lattice models},
  arXiv:hep-th/9311037.

  L.~D.~Faddeev and G.~P.~Korchemsky,
  {\em High-energy QCD as a completely integrable model},
  Phys.\ Lett.\  B {\bf 342}, 311 (1995)
  [arXiv:hep-th/9404173].

  G.~P.~Korchemsky,
  {\em Bethe Ansatz For QCD Pomeron},
  Nucl.\ Phys.\  B {\bf 443}, 255 (1995)
  [arXiv:hep-ph/9501232].

  V.~M.~Braun, S.~E.~Derkachov and A.~N.~Manashov,
  {\em Integrability of three-particle evolution equations in {QCD}},
  Phys.\ Rev.\ Lett.\  {\bf 81}, 2020 (1998)
  [arXiv:hep-ph/9805225].

\bibitem{Belitsky:2004sc}
  A.~V.~Belitsky, S.~E.~Derkachov, G.~P.~Korchemsky and A.~N.~Manashov,
  {\em Dilatation operator in (super-)Yang-Mills theories on the light-cone},
  Nucl.\ Phys.\  B {\bf 708}, 115 (2005)
  [arXiv:hep-th/0409120].

\bibitem{Belitsky:2004sf}
  A.~V.~Belitsky, G.~P.~Korchemsky and D.~Mueller,
  {\em Integrability in Yang-Mills theory on the light cone beyond leading order},
  Phys.\ Rev.\ Lett.\  {\bf 94}, 151603 (2005)
  [arXiv:hep-th/0412054].

\bibitem{Belitsky:2005bu}
  A.~V.~Belitsky, G.~P.~Korchemsky and D.~Mueller,
  {\em Integrability of two-loop dilatation operator in gauge theories},
  Nucl.\ Phys.\  B {\bf 735}, 17 (2006)
  [arXiv:hep-th/0509121].
  
\bibitem{DiVecchia:2004jw}
  P.~Di Vecchia and A.~Tanzini,
  {\em N = 2 super Yang-Mills and the XXZ spin chain},
  J.\ Geom.\ Phys.\  {\bf 54}, 116 (2005)
  [arXiv:hep-th/0405262].

\bibitem{Belitsky:2005gr}
  A.~V.~Belitsky, S.~E.~Derkachov, G.~P.~Korchemsky and A.~N.~Manashov,
  {\em Superconformal operators in Yang-Mills theories on the light-cone},
  Nucl.\ Phys.\  B {\bf 722}, 191 (2005)
  [arXiv:hep-th/0503137].

\bibitem{Wang:2003cu}
  X.~J.~Wang and Y.~S.~Wu,
  {\em Integrable spin chain and operator mixing in N = 1,2 supersymmetric theories},
  Nucl.\ Phys.\  B {\bf 683}, 363 (2004)
  [arXiv:hep-th/0311073]. See also  N.~Beisert and R.~Roiban, ``The Bethe ansatz for Z(S) orbifolds of N = 4 super Yang-Mills theory,'' JHEP {\bf 0511}, 037 (2005)
  [arXiv:hep-th/0510209] 
  
  D.~Astolfi, V.~Forini, G.~Grignani and G.~W.~Semenoff,
  ``Finite size corrections and integrability of N = 2 SYM and DLCQ strings on
  a pp-wave,''
  JHEP {\bf 0609}, 056 (2006)
  [arXiv:hep-th/0606193].

\bibitem{Maldacena:1997re}
  J.~M.~Maldacena,
  {\em The large $N$ limit of superconformal field theories and supergravity},
  Adv.\ Theor.\ Math.\ Phys.\  {\bf 2}, 231 (1998)
  [Int.\ J.\ Theor.\ Phys.\  {\bf 38}, 1113 (1999)]
  [arXiv:hep-th/9711200].

  S.~S.~Gubser, I.~R.~Klebanov and A.~M.~Polyakov,
  {\em Gauge theory correlators from non-critical string theory},
  Phys.\ Lett.\ B {\bf 428}, 105 (1998)
  [arXiv:hep-th/9802109].

  E.~Witten,
  {\em Anti-de Sitter space and holography},
  Adv.\ Theor.\ Math.\ Phys.\  {\bf 2}, 253 (1998)
  [arXiv:hep-th/9802150].


  I.~R.~Klebanov,
  {\em TASI lectures: Introduction to the AdS/CFT correspondence},
  Lectures given at Theoretical Advanced Study Institute in Elementary Particle Physics (TASI 99): Strings, Branes, and Gravity, Boulder, Colorado, 31 May - 25 Jun 1999.
  Published in ``Boulder 1999, Strings, branes and gravity'', 615-650 ,
  arXiv:hep-th/0009139.
  
  
\bibitem{Beisert:2004ry}
  N.~Beisert,
  {\em ``The dilatation operator of N = 4 super Yang-Mills theory and
    integrability},
  Phys.\ Rept.\  {\bf 405} (2005) 1,
  [arXiv:hep-th/0407277].
    
\bibitem{Plefka:2005bk}
  J.~Plefka,
{\em  Spinning strings and integrable spin chains in the AdS/CFT
  correspondence},
  arXiv:hep-th/0507136.


\bibitem{Bena:2003wd}
  I.~Bena, J.~Polchinski and R.~Roiban,
  {\em Hidden symmetries of the $\ads$ superstring},
  Phys.\ Rev.\  D {\bf 69}, 046002 (2004)
  [arXiv:hep-th/0305116].


\bibitem{Arutyunov:2004vx}
  G.~Arutyunov, S.~Frolov and M.~Staudacher,
  {\em Bethe ansatz for quantum strings},
  JHEP {\bf 0410}, 016 (2004)
  [arXiv:hep-th/0406256].


\bibitem{Staudacher:2004tk}
  M.~Staudacher,
  {\em The factorized S-matrix of CFT/AdS},
  JHEP {\bf 0505}, 054 (2005)
  [arXiv:hep-th/0412188].
  
\bibitem{Park:2005ji}
  I.~Y.~Park, A.~Tirziu and A.~A.~Tseytlin,
  ``Spinning strings in AdS(5) x S**5: One-loop correction to energy in  SL(2)
  sector,''
  JHEP {\bf 0503}, 013 (2005)
  [arXiv:hep-th/0501203].

\bibitem{Beisert:2005fw}
  N.~Beisert and M.~Staudacher,
  {\em Long-range PSU(2,2|4) Bethe ansaetze for gauge theory and strings},
  Nucl.\ Phys.\  B {\bf 727}, 1 (2005)
  [arXiv:hep-th/0504190].

\bibitem{Beisert:2005tm}
  N.~Beisert,
  {\em The su(2|2) dynamic S-matrix},
  arXiv:hep-th/0511082.


\bibitem{Janik:2006dc}
  R.~A.~Janik,
  {\em The $\ads$ superstring worldsheet S-matrix and crossing  symmetry},
  Phys.\ Rev.\  D {\bf 73}, 086006 (2006)
  [arXiv:hep-th/0603038].
  
\bibitem{Eden:2006rx}
  B.~Eden and M.~Staudacher,
  {\em Integrability and transcendentality},
  J.\ Stat.\ Mech.\  {\bf 0611}, P014 (2006)
  [arXiv:hep-th/0603157].
  
\bibitem{Hernandez:2006tk}
  R.~Hernandez and E.~Lopez,
  {\em Quantum corrections to the string Bethe ansatz},
  JHEP {\bf 0607}, 004 (2006)
  [arXiv:hep-th/0603204].
  
  \bibitem{Freyhult:2006vr}
  L.~Freyhult and C.~Kristjansen,
  ``A universality test of the quantum string Bethe ansatz,''
  Phys.\ Lett.\  B {\bf 638}, 258 (2006)
  [arXiv:hep-th/0604069].

\bibitem{Plefka:2006ze}
  J.~Plefka, F.~Spill and A.~Torrielli,
 {\em On the Hopf algebra structure of the AdS/CFT S-matrix},
  Phys.\ Rev.\  D {\bf 74}, 066008 (2006)
  [arXiv:hep-th/0608038].


\bibitem{Beisert:2006ib}
  N.~Beisert, R.~Hernandez and E.~Lopez,
  {\em A crossing-symmetric phase for $\ads$ strings},
  JHEP {\bf 0611}, 070 (2006)
  [arXiv:hep-th/0609044].


\bibitem{Beisert:2006ez}
  N.~Beisert, B.~Eden and M.~Staudacher,
  {\em Transcendentality and crossing},
  J.\ Stat.\ Mech.\  {\bf 0701}, P021 (2007)
  [arXiv:hep-th/0610251].




\bibitem{Dokshitzer:2006nm}
  Yu.~L.~Dokshitzer and G.~Marchesini,
  {\em N = 4 SUSY Yang-Mills: Three loops made simple(r)},
  Phys.\ Lett.\  B {\bf 646}, 189 (2007)
  [arXiv:hep-th/0612248].

%


\bibitem{Kotikov:2007cy}
  A.~V.~Kotikov, L.~N.~Lipatov, A.~Rej, M.~Staudacher and V.~N.~Velizhanin,
  {\em Dressing and Wrapping},
  arXiv:0704.3586 [hep-th].



\bibitem{Janik:2007wt}
  R.~A.~Janik and T.~Lukowski,
  {\em Wrapping interactions at strong coupling -- the giant magnon},
  arXiv:0708.2208 [hep-th].

\bibitem{Rej:2007vm}
  A.~Rej, M.~Staudacher and S.~Zieme,
  {\it Nesting and dressing},
  arXiv:hep-th/0702151.

\bibitem{Lipatov:1997vu}
  L.~N.~Lipatov,
   {\it "Evolution equations in QCD",}
     in ``Perspectives in Hadronic Physics,''
     Proceedings of the Conference, ICTP, Trieste, Italy, 12-16 May 1997,
     eds.~S.~Boffi, C.~Ciofi Degli Atti and M.~Giannini,
     World Scientific (Singapore, 1998).

\bibitem{Lipatov:1976zz}
  L.~N.~Lipatov,
  {\it ``Reggeization of the vector meson and the
  vacuum singularity in nonabelian gauge theories,''}
  Sov.\ J.\ Nucl.\ Phys.\  {\bf 23} (1976) 338
  [Yad.\ Fiz.\  {\bf 23} (1976) 642].

  E.~A.~Kuraev, L.~N.~Lipatov and V.~S.~Fadin,
  {\it ``The Pomeranchuk singularity in nonabelian gauge theories,''}
  Sov.\ Phys.\ JETP {\bf 45} (1977) 199
  [Zh.\ Eksp.\ Teor.\ Fiz.\  {\bf 72} (1977) 377].

  I.~I.~Balitsky and L.~N.~Lipatov,
  {\it ``The Pomeranchuk singularity in Quantum Chromodynamics,''}
  Sov.\ J.\ Nucl.\ Phys.\  {\bf 28} (1978) 822
  [Yad.\ Fiz.\  {\bf 28} (1978) 1597].

\bibitem{twist}
  M.~Beccaria,
  {\em Three loop anomalous dimensions of twist-3 gauge operators in N=4 SYM},
  arXiv:0707.1574 [hep-th].

  M.~Beccaria, Yu.~L.~Dokshitzer and G.~Marchesini,
  {\em Twist 3 of the sl(2) sector of N=4 SYM and reciprocity respecting evolution},
  Phys.\ Lett.\  B {\bf 652}, 194 (2007)
  [arXiv:0705.2639 [hep-th]].

  M.~Beccaria,
  {\em Universality of three gaugino anomalous dimensions in N = 4 SYM},
  JHEP {\bf 0706}, 054 (2007)
  [arXiv:0705.0663 [hep-th]].

  M.~Beccaria,
  {\em Anomalous dimensions at twist-3 in the sl(2) sector of N = 4 SYM},
  JHEP {\bf 0706}, 044 (2007)
  [arXiv:0704.3570 [hep-th]].

\bibitem{Lubcke:2004dg}
  M.~Lubcke and K.~Zarembo,
  {\em Finite-size corrections to anomalous dimensions in N = 4 SYM theory},
  JHEP {\bf 0405}, 049 (2004)
  [arXiv:hep-th/0405055].

\bibitem{Beisert:2005bv}
  N.~Beisert and L.~Freyhult,
  {\em Fluctuations and energy shifts in the Bethe ansatz},
  Phys.\ Lett.\  B {\bf 622}, 343 (2005)
  [arXiv:hep-th/0506243].

\bibitem{Gromov:2005gp}
  N.~Gromov and V.~Kazakov,
  {\em Double scaling and finite size corrections in sl(2) spin chain},
  Nucl.\ Phys.\  B {\bf 736}, 199 (2006)
  [arXiv:hep-th/0510194].

\bibitem{Feverati:2006hh}
  G.~Feverati, D.~Fioravanti, P.~Grinza and M.~Rossi,
  {\em Hubbard's adventures in N = 4 SYM-land? Some non-perturbative considerations on finite length operators},
  J.\ Stat.\ Mech.\  {\bf 0702}, P001 (2007)
  [arXiv:hep-th/0611186].

  G.~Feverati, D.~Fioravanti, P.~Grinza and M.~Rossi,
  {\em On the finite size corrections of anti-ferromagnetic anomalous  dimensions in N = 4 SYM},
  JHEP {\bf 0605}, 068 (2006)
  [arXiv:hep-th/0602189].
  
  
\bibitem{Casteill:2007ct}
  P.~Y.~Casteill and C.~Kristjansen,
  ``The Strong Coupling Limit of the Scaling Function from the Quantum   String
  Bethe Ansatz,''
  Nucl.\ Phys.\  B {\bf 785}, 1 (2007)
  [arXiv:0705.0890 [hep-th]].
  
  
\bibitem{Freyhult:2004iq}
  L.~Freyhult,
  ``Bethe ansatz and fluctuations in SU(3) Yang-Mills operators,''
  JHEP {\bf 0406}, 010 (2004)
  [arXiv:hep-th/0405167].
  
\bibitem{Freyhult:2005fn}
  L.~Freyhult and C.~Kristjansen,
  ``Finite size corrections to three-spin string duals,''
  JHEP {\bf 0505}, 043 (2005)
  [arXiv:hep-th/0502122].

\bibitem{deVega:1987vh}
  H.~J.~de Vega,
  {\em Finite Size Corrections For Nested Bethe Ansatz Models And Conformal Invariance},
  J.\ Phys.\ A  {\bf 20}, 6023 (1987).
  
\bibitem{Gromov:2007ky}
  N.~A.~Gromov and P.~Vieira,
  {\em Complete 1-loop test of AdS/CFT},
  arXiv:0709.3487 [hep-th].
  
  \bibitem{Kazakov:2004qf}
  V.~A.~Kazakov, A.~Marshakov, J.~A.~Minahan and K.~Zarembo,
  {\em Classical / quantum integrability in AdS/CFT},
  JHEP {\bf 0405}, 024 (2004)
  [arXiv:hep-th/0402207].
  
  
\bibitem{Beisert:2005bm}
  N.~Beisert, V.~A.~Kazakov, K.~Sakai and K.~Zarembo,
  ``The algebraic curve of classical superstrings on AdS(5) x S**5,''
  Commun.\ Math.\ Phys.\  {\bf 263}, 659 (2006)
  [arXiv:hep-th/0502226].
  
\bibitem{Beisert:2005cw}
  N.~Beisert and A.~A.~Tseytlin,
  ``On quantum corrections to spinning strings and Bethe equations,''
  Phys.\ Lett.\  B {\bf 629}, 102 (2005)
  [arXiv:hep-th/0509084].

\bibitem{SchaferNameki:2006ey}
  S.~Schafer-Nameki, M.~Zamaklar and K.~Zarembo,
  ``How accurate is the quantum string Bethe ansatz?,''
  JHEP {\bf 0612}, 020 (2006)
  [arXiv:hep-th/0610250].
  

\bibitem{Ambjorn:2005wa}
  J.~Ambjorn, R.~A.~Janik and C.~Kristjansen,
  ``Wrapping interactions and a new source of corrections to the spin-chain  /
  string duality,''
  Nucl.\ Phys.\  B {\bf 736}, 288 (2006)
  [arXiv:hep-th/0510171].
  
\bibitem{SchaferNameki:2006gk}
  S.~Schafer-Nameki,
  ``Exact expressions for quantum corrections to spinning strings,''
  Phys.\ Lett.\  B {\bf 639}, 571 (2006)
  [arXiv:hep-th/0602214].
  
\bibitem{Hofman:2006xt}
  D.~M.~Hofman and J.~M.~Maldacena,
  ``Giant magnons,''
  J.\ Phys.\ A  {\bf 39}, 13095 (2006)
  [arXiv:hep-th/0604135].
  
\bibitem{Arutyunov:2006gs}
  G.~Arutyunov, S.~Frolov and M.~Zamaklar,
  ``Finite-size effects from giant magnons,''
  Nucl.\ Phys.\  B {\bf 778}, 1 (2007)
  [arXiv:hep-th/0606126].
  
\bibitem{Astolfi:2007uz}
  D.~Astolfi, V.~Forini, G.~Grignani and G.~W.~Semenoff,
  ``Gauge invariant finite size spectrum of the giant magnon,''
  Phys.\ Lett.\  B {\bf 651}, 329 (2007)
  [arXiv:hep-th/0702043].
  
\bibitem{Ferretti:2004ba}
  G.~Ferretti, R.~Heise and K.~Zarembo,
  {\it ``New integrable structures in large-N QCD,''}
  Phys.\ Rev.\  D {\bf 70} (2004) 074024,
  [arXiv:hep-th/0404187].

\bibitem{Beisert:2004fv}
  N.~Beisert, G.~Ferretti, R.~Heise and K.~Zarembo,
  {\it ``One-loop QCD spin chain and its spectrum,''}
  Nucl.\ Phys.\  B {\bf 717} (2005) 137,
  [arXiv:hep-th/0412029].

\bibitem{SchaferNameki:2004ik}
  S.~Schafer-Nameki,
  {\em The algebraic curve of 1-loop planar N = 4 SYM},
  Nucl.\ Phys.\  B {\bf 714}, 3 (2005)
  [arXiv:hep-th/0412254].
  

\bibitem{Beisert:2003jj}
  N.~Beisert,
  {\em The complete one-loop dilatation operator of N = 4 super Yang-Mills theory},
  Nucl.\ Phys.\  B {\bf 676}, 3 (2004)
  [arXiv:hep-th/0307015].

\bibitem{Gorsky:2004kb}
  A.~Gorsky,
  {\em Gauge / string duality: First achievements},
  8th International Moscow School of Physics and 33rd ITEP Winter School of Physics, Moscow, Russia, 22 Feb - 2 Mar 2005,
  Surveys High Energ.\ Phys.\  {\bf 19}, 233 (2004).

\bibitem{'tHooft:1976fv}
  G.~'t Hooft,
  {\em Computation of the quantum effects due to a four-dimensional pseudoparticle},
  Phys.\ Rev.\  D {\bf 14}, 3432 (1976)
  [Erratum-ibid.\  D {\bf 18}, 2199 (1978)].


\bibitem{spin1}
  A.~B.~Zamolodchikov and V.~A.~Fateev,
  {\em Model Factorized S Matrix And An Integrable Heisenberg Chain With Spin 1. (In Russian)},
  Yad.\ Fiz.\  {\bf 32} (1980) 581.

  P.~P.~Kulish, N.~Y.~Reshetikhin and E.~K.~Sklyanin,
  {\em Yang-Baxter Equation And Representation Theory. 1},
  Lett.\ Math.\ Phys.\  {\bf 5} (1981) 393.

  N.~Y.~Reshetikhin,
  {\em Integrable models of quantum one-dimensional magnets with $O(K)$ and $Sp(2K)$ symmetry},
  Theor.\ Math.\ Phys.\  {\bf 63} (1985) 555
  [Teor.\ Mat.\ Fiz.\  {\bf 63} (1985) 347].

\bibitem{Frolov:2003qc}
  S.~Frolov and A.~A.~Tseytlin,
 ``Multi-spin string solutions in AdS(5) x S**5,''
  Nucl.\ Phys.\  B {\bf 668}, 77 (2003)
  [arXiv:hep-th/0304255].

\bibitem{Park:2005kt}
  I.~Y.~Park, A.~Tirziu and A.~A.~Tseytlin,
  {\em Semiclassical circular strings in AdS(5) and 'long' gauge field  strength operators},
  Phys.\ Rev.\  D {\bf 71}, 126008 (2005)
  [arXiv:hep-th/0505130].


\bibitem{Beisert:2005di}
  N.~Beisert, V.~A.~Kazakov, K.~Sakai and K.~Zarembo,
  {\em Complete spectrum of long operators in N = 4 SYM at one loop},
  JHEP {\bf 0507}, 030 (2005)
  [arXiv:hep-th/0503200].



\bibitem{Beisert:2007hz}
  N.~Beisert, T.~McLoughlin and R.~Roiban,
  {\em The Four-Loop Dressing Phase of N=4 SYM},
  Phys.\ Rev.\  D {\bf 76}, 046002 (2007)
  [arXiv:0705.0321 [hep-th]].

\end{thebibliography}
\end{document}